# Boosting advice and knowledge sharing among healthcare professionals

Fronzetti Colladon, A., Grippa, F., Broccatelli, C., Mauren, C., Mckinsey, S., Kattan, J., Sutton, E. S. J., Satlin, L., & Bucuvalas, J.

# Boosting advice and knowledge sharing among healthcare professionals

*Fronzetti Colladon, A., Grippa, F., Broccatelli, C., Mauren, C., Mckinsey, S., Kattan, J., Sutton, E. S. J., Satlin, L., & Bucuvalas, J.*

**Abstract**

**Purpose:** This study investigates the dynamics of knowledge sharing in healthcare, exploring some of the factors that are more likely to influence the evolution of idea sharing and advice seeking in healthcare.

**Design/methodology/approach:** We engaged 50 pediatricians representing many subspecialties at a mid-size US children's hospital using a social network survey to map and measure advice seeking and idea sharing networks. Through the application of Stochastic Actor-Oriented Models, we compared the structure of the two networks prior to a leadership program and eight weeks post conclusion.

**Findings:** Our models indicate that healthcare professionals carefully and intentionally choose with whom they share ideas and from whom to seek advice. The process is fluid, non-hierarchical and open to changing partners. Significant transitivity effects indicate that the processes of knowledge sharing can be supported by mediation and brokerage.

**Originality:** Hospital administrators can use this method to assess knowledge-sharing dynamics, design and evaluate professional development initiatives, and promote new organizational structures that break down communication silos. Our work contributes to the literature on knowledge sharing in healthcare by adopting a social network approach, going beyond the dyadic level, and assessing the indirect influence of peers' relationships on individual networks.

1. Introduction

The healthcare community is increasingly recognizing the need to find new approaches to improve outcomes and the overall experience for patients and healthcare workers, which requires innovative ways to break down isolated professional silos and tribes (Alrahbi *et al.*, 2022; Antonacci *et al.*, 2017; Long *et al.*, 2013; Seid *et al.*, 2014). While traditional diagnostic decision making is often associated with the responsibility of an individual clinician, the increased complexity of care calls for a more holistic approach to medicine, where a complete diagnosis and cure requires frequent interaction among healthcare professionals from different disciplines with different cultures, skills, connections and specialized knowledge (McKee, 1988; Ventegodt *et al.*, 2003). To solve complex clinical cases, healthcare professionals need to build ties across units within and outside the organization and flexibly adapt their membership and leadership structure over time to incorporate expertise that contributes to decision-making (Ancona *et al.*, 2002; Rios-Ballesteros and Fuerst, 2021). As discussed in the open innovation literature, to accelerate internal innovation it is key to explore and exploit connections within and across organizational boundaries, using inflows and outflows of knowledge (Chesbrough *et al.*, 2006; Gassmann *et al.*, 2010). As the distinction between members and non-members becomes fluid, knowledge flow within and across them becomes increasingly dispersed (Mortensen and Haas, 2018). Each team member brings a unique set of special knowledge and capabilities, as well as interpersonal relationships with key stakeholders. Today, dispersed individuals are engaged in collecting ideas and advance knowledge creation and diffusion through mechanisms that break down institutional barriers (Bican *et al.*, 2017; Cappa, 2022). Multiple studies have demonstrated that working in cross-functional teams has the potential to reduce errors as healthcare professionals rely on each other's expertise and specialized knowledge to complete complex tasks (Delva *et al.*, 2008; Lemieux-Charles and McGuire, 2006; Palazzolo *et al.*, 2011). Healthcare interdisciplinary teams have proven successful in improving communication and knowledge exchange across hospital units, reducing the multiple gaps that exist among professions, departments, and specialties, including the clinician-patient divide (Grippa *et al.*, 2018; Pereira de Souza *et al.*, 2021; Zhang and Wang, 2021).

Despite the wide recognition that knowledge sharing and looking for outside perspectives is key to solving complex healthcare tasks, evidence shows that working in cross-disciplinary units or teams without adequate support or access to professional networks may lead to negative outcomes (Gilardi *et al.*, 2014; Willard-Grace *et al.*, 2014). To facilitate knowledge integration, especially when faced with complex new cases, healthcare professionals need to be equipped with the managerial skills and network ties to identify and confront the dependencies across the knowledge boundaries (Majchrzak *et al.*, 2012).

Most of the studies thus far have explored advice seeking and knowledge sharing through the lens of individual factors, including individual anxiety (Gino *et al.*, 2012), the advisors' perceived expertise (Yaniv and Milyavsky, 2007), trust and accessibility (Hofmann *et al.*, 2009) or the ability to access knowledge at no cost (Gino, 2008). In this exploratory study, we adopt a network science approach to explore such questions as: what are the network properties that may influence the individual ability to share knowledge? What are the network effects that may influence the ability of healthcare professionals to reach out for help?

To address these questions, we conducted a social network analysis before and after a leadership program at a large urban academic medical center. The training program involved pediatricians recognized as experts in their subspecialties who were not in a formally defined system or divisional leadership role. The goal of the leadership program was to facilitate the creation of interdisciplinary ties and decrease barriers to collaboration. Our study contributes to knowledge creation and advice seeking literature in healthcare by describing how the structural properties of the advice seeking and idea sharing networks evolve after a training program. This study provides practical insights to design strategies that seek to promote a balanced exchange of advice and idea sharing among healthcare professionals.

The paper is organized as follows. In section 2, we discuss the literature on knowledge sharing and advice seeking with a focus on healthcare. In section 3, we illustrate the research setting, the methods of data collection and present the Stochastic Actor-Oriented Models. In section 4, we present the results through a social network representation of pre- and post-leadership connections, as well as the effects for modeling social dynamics. Finally, in Section 5 we describe both theoretical and practical implications of the study, concluding with future directions and study limitations.

## 2. Measuring Idea Sharing and Advice Seeking

A growing body of literature known as knowledge networks research has highlighted how social relationships impact the efficacy and efficiency by which individuals, teams, and organizations create knowledge by influencing their ability to access, transfer, absorb, and apply knowledge (Phelps *et al.*, 2012; Tsai, 2001; Ward *et al.*, 2012). Weak ties are relevant for accessing new information and receiving valuable feedback (Granovetter, 1973), while diverse and heterogeneous contacts generate learning, increase resilience and build innovation opportunities through indirect bridging and linking ties (Aldrich and Meyer, 2015).

An organization's ability to perform well through collective decision-making depends on its internal and external ties (Ancona *et al.*, 2002) and its ability to explore new information and solutions. Creative performance is mediated by the ability to share different types of knowledge, such as information and know-how (Kessel *et al.*, 2012). Effective idea sharing and knowledge creation depend on the ability to investigate new solutions by leveraging individual advice networks (Shore *et al.*, 2015). Individual innovative behaviors are not isolated but actions embedded in the interconnected organizational fabric (Cangialosi *et al.*, 2021). When individuals reach out to others for work-related advice, there is more than an answer and a solution that flows between them. A study in the healthcare environment investigating the sensemaking processes underlying how nurses decide whom to ask for advice found that nurses are more likely to seek help from peers they perceive as experts, if they perceive them as accessible, trustworthy, or both (Hofmann *et al.*, 2009).

Cross and colleagues (2001) explored five different dimensions of the advice network and illustrated the benefits for individuals to seek information through their network of contacts. Seeking advice provides people with the following benefits: alternative solutions to problems; meta-knowledge; problem reformulation (thinking differently about a specific problem); validation; and legitimation. Identifying information brokers and measuring the transfer of tacit knowledge and expertise have been identified as important challenges in healthcare (Waring *et al.*, 2013).

Many studies have focused on team collaboration by investigating the composition of teams using traditional social science methodologies. Others have focused on the role of individual attributes, such as personality types or behavioral patterns (Srinivasan and Uddin, 2017). For instance, Kilpatrick (2013) describes the impact of introducing the intermediary role of acute care nurse practitioner on decision making and communication. Our study contributes to the understanding of the factors facilitating knowledge sharing by adopting a social network perspective to map the connections developed among medical professionals. (Ripley *et al.*, 2020; Wasserman and Faust, 1994). Recently, an increasing number of studies in healthcare settings have been using social network analysis to identify information brokers who span team and organizational boundaries and identify networks' fragmentation points (Griffiths *et al.*, 2012; Srinivasan and Uddin, 2017). For example, a longitudinal email network analysis involving two chronic care teams at a US children's hospital demonstrated how improving awareness of team communication and relying on information brokers can lead to increased density and inter-role connectedness (Grippa *et al.*, 2011; Palazzolo *et al.*, 2011). Another study exploring collaboration networks among Australian medical providers found that network features like degree, brokerage, closeness, and the number of triadic relationships have a statistically significant impact on efficiency metrics – such as medical costs, length of stay, and complication rate (Wang *et al.*, 2014). Other studies have used social network analysis to illustrate how healthcare teams differentiate their professional connections through a selection of different colleagues when asking for advice on clinical practice and when choosing people to share professional development ideas (Burt *et al.*, 2012). Similarly, our study adopts a social network perspective. Differently from them, we analyze the properties of teams sharing ideas and seeking advice before and after a specific initiative.

## 3. Research Methodology

To explore the cross-network influence between advice seeking and idea generation networks, we mapped the cross-division interactions developed before and after a leadership program catered to physicians in a children's hospital. The cohort was composed of approximately 50 faculty members working at different medical campuses in New York City and representing various specialties, from newborn medicine to cardiology, gastroenterology, allergy, and general pediatrics. This represented an optimal research setting for our study, given the interdisciplinary nature of the work conducted at these medical facilities by pediatricians who longed for mechanisms to bridge the physical distance between colleagues. The traditional target for these learning opportunities had always been senior faculty and rarely included faculty leading front-line clinical teams. Allowing physicians from all levels of the organization to join is aligned with the concept of distributed leadership, stating that leadership should rest with whoever is best positioned to exercise it, regardless of title (Ancona *et al.*, 2007). It is worth noting that our participants did not have hierarchical dependencies that may affect their interaction dynamics and were all from different work teams. They were sometimes from the same division (a control variable we include in our models). While we could not collect individual data regarding tenure, we were reassured that participants had been working at the hospital for at least two years. This was useful to exclude the possibility of individuals having few connections due to being newcomers.

Participants volunteered to participate in a 16-week modular leadership program, with each module containing lectures from speakers of a top Business School in the US, case studies, and team assignments. The goals of the Leadership program were to develop the mindset and behaviors essential to thrive in a Volatile, Uncertain, Complex, and Ambiguous (VUCA) environment; demonstrate resilient, adaptive responses to organizational and leadership challenges; and create a more agile, flexible, and change-ready culture that enables a rapid and focused response to VUCA changes. The modules included topics like thriving in a VUCA environment, leading with impact, and collaborating for results. The program was designed in a format to encourage cross-disciplinary team formation. Participants voluntarily joined the leadership program, and as part of the screening process, they shared career development goals and expectations on the program's impact on their careers.

### 3.1. Data Collection

Participants completed two surveys to map their social networks related to advice seeking and idea sharing. The first survey mapped their networks prior to the leadership program, while the second provided network maps after eight weeks from the course completion. Outside of this leadership program, respondents worked in separate hospital divisions and often crossed units in interdisciplinary teams based on the needs of patients. These pediatricians were at the same level within the organization and did not report to each other. They often shared patients depending on the complexity of the cases, which required cross-disciplinary interventions. Within the leadership programs, they were organized in teams for the purpose of the educational experience.

Because many of the participants were not direct colleagues outside of the leadership program, we framed the questions in a general and comprehensive way. The first question was associated with advice seeking regarding a patient situation (*Imagine you are faced with a tough patient-related problem: from whom do you seek advice?*). The question could receive more than one answer. The second question was more general, as it aimed to map any colleagues worth sharing ideas with ("*Imagine you come up with a new idea to solve an existing problem: who are the colleagues with whom you would you share your ideas?*"). We had 47 respondents for the pre-course survey and 42 for the post-course survey. The final social networks of respondents, as well as the people reported by others, included a total of 233 individuals pre-course and 252 individuals post-course, holding various roles from nurses and physicians (68%) to administration (4.7%), leadership (15%), research (4.7%) and other roles (6.8%). We also collected data on participant gender and hospital division.

### 3.2. Stochastic Actor-Oriented Models

To understand which factors and dynamics could influence knowledge sharing relationships, we used Stochastic Actor-Oriented Models (SAOMs) (Snijders *et al.*, 2010; Steglich *et al.*, 2010). Here we provide a general overview of these models as explained by Ripley and colleagues (2020) in relation to this specific work. SAOMs are actor-oriented models where network ties depend on the actor's choices to create, maintain, or modify their ties over time, assuming they have the chance to do so. Individual choices explain how ties are distributed within a network and how individual behavior can be influenced by someone's ties with others. Unlike traditional statistical methods like regression analysis, SAOMs can simultaneously handle the evolution of network dynamics and behaviors. They do that by modeling actor choices in mini-steps, counting the possible changes that can occur across different time points in sequence. The models theorize actors' tie changes as parameters that express individual tendencies to form (or eliminate) interactions embedded in network structures. These can also represent changes in behaviors (influence) as well as controlling for individual's preferences in selecting their partners based on their characteristics (selection). Our models are based on some of the most recent developments in the study of SAOMs (Aronson, 2016; Block, 2015).

We applied SAOMs to investigate how idea sharing ties affect the evolution of advice seeking ties and vice versa. We also considered how individual membership in different hospital divisions might influence the formation of ties in dyadic or triadic structures. In particular, we used the SOAM extension for multivariate networks that model the evolution of two networks simultaneously (Elmer *et al.*, 2017; Snijders *et al.*, 2013, 2020; Stadtfeld and Pentland, 2015). We estimated SAOMs using the "RSiena" (Simulation Investigation for Empirical Network Analyses) library, available in R statistical software (Ripley *et al.*, 2020). Our networks include information on the directionality of ties and contain 252 individuals based on the combination of all individuals present at both points in time (before and after the leadership program). Participants present in only one of the two observations were treated as structural zeros, meaning they were not able to send and receive ties during the time of the observation for which they were missing. In this way, changes among these nodes are not possible, and thus they are excluded from the simulations (Ripley *et al.*, 2020). Finally, the statistical approach used here can serve to understand some potential dynamics that are likely to explain the presence (or absence) of ties within the given network, whose boundaries are defined by the workshop attendees. A different network design would be required to collect data representing the whole population of health professionals at the hospital.

In the models, we refer to commonly used terms in network studies, including: network density, i.e., the ratio of actual and potential network connections; node degree, i.e., the number of different connections a person has (e.g., advice sharing relationships). We also differentiate between incoming and outgoing arcs, using the terms of in- and out-degree, respectively. For example, if actor A seeks actor B for advice, we find an arc starting at A and terminating at B in the advice network (Wasserman and Faust, 1994). The analysis also considers structural effects such as reciprocity and transitivity. Reciprocity is the individuals' tendency to reciprocate ties, while transitivity is the tendency to form triads. This depicts the situation where my advisors' advisors also become my advisors. Effects depending on the degree, the outgoing and incoming ties (so-called out-degree activity and indegree popularity effects) are also included. These effects are considered endogenous because they refer to network structures. Instead, membership by division is considered as actor covariate and represents an exogenous variable since it is a given individual attribute. We used this variable to explore the preference to select advisors and share ideas with members of the same division in dyads and triads (also called homophily tendency). Finally, we included activity, popularity, and several triadic cross-network effects that investigate the tendencies to create, maintain or remove ties in one network based on the presence of ties in the other network. Table 1 illustrates the complete list of effects we used in our models and the corresponding visual representation and interpretation.

| Effect | RSiena parameter name | Visualization | Description (if positive and significant) |
|---|---|---|---|
| basic rate parameter | rate | NA | Frequency in seeking advice and sharing ideas. |
| outdegree (density) | density | $t_1$ $t_2$ — advice ⋯⋯> ideas sharing | Basic tendency to seek advice/share ideas at all (intercept). |

| reciprocity | recip | 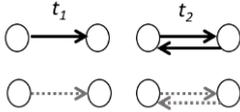 | Individuals tend to give each other advice and to share ideas reciprocally. |
|---|---|---|---|
| transitive reciprocated triplets | transRecTrip | 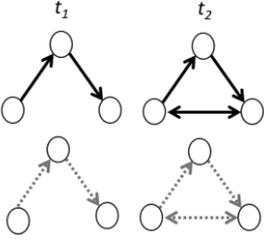 | Individuals tend to close two-paths through a reciprocated tie. |
| transitivity (gwesp) | gwespFF | 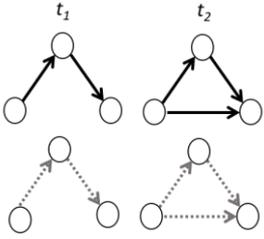 | Individuals tend to close indirect connections and to tie with the colleague of their colleague. |
| indegree - popularity (sqrt) | inPopSqrt | 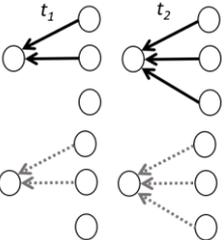 | Popular individuals tend to attract advice seekers/ receive other's people ideas even more ("Matthew effect" or preferential attachment). |
| outdegree - activity (centered) | outAct.c | 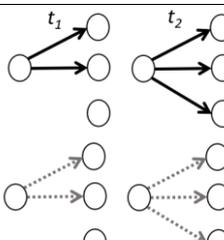 | Individuals actively seeking for advice/ share ideas tend to seek advice/ share ideas even more. |
| positive outdegree effect | outTrunc(1) | NA | Tendency to be an isolate with respect to outgoing ties or have out-degree 1. |
| anti in-isolates | antiInIso | NA | Individuals tend to connect to others otherwise isolated, i.e., without incoming or outgoing ties. |
| homophily - same division | sameX | 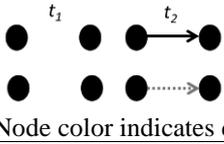<br>Node color indicates division. | Individuals tend to connect with others who belong to their same division. |
| transitive triplets jumping by division | jumpXTransTrip | 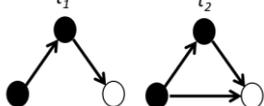 | Individuals tend to seek advice outside their division if their advisors act as intermediaries. |

| mix degree effect - indegree popularity (sqrt) | inPopIntn | $t_1$ $t_2$ | Popular advisors also tend to be attractive for ideas sharing. Popular individuals that receive many ideas also tend to be popular advisors. |
|---|---|---|---|
| mix degree effect - outdegree activity (sqrt) | outActIntn | $t_1$ $t_2$ | Individuals who seek many advices also tend to share many ideas. Individuals who share many ideas, also tend to seek for many advices. |
| advice: closure of ideas | closure | $t_1$ $t_2$ | Individuals tend to close indirect connections of shared ideas and to share advice with the colleague of their colleague. |
| advice: shared incoming ideas | sharedIn | $t_1$ $t_2$ | Individuals who receive ideas from the same third person tend to share advice. |
| ideas: from advice agreement | from | $t_1$ $t_2$ | Individuals that have an advisor in common tend to share ideas. |

**Table 1.** Effects for modeling social dynamics.

## 4. Results

Figure 1 shows the changes in relationships before (T1) and after the leadership course (T2). Hospital professionals are represented as nodes, while advice- and idea sharing ties are represented as arcs. Networks on the left correspond to the pre-course sharing of advice and ideas (T1); networks on the right represent the exchanges of advice and ideas after the conclusion of the course (T2).

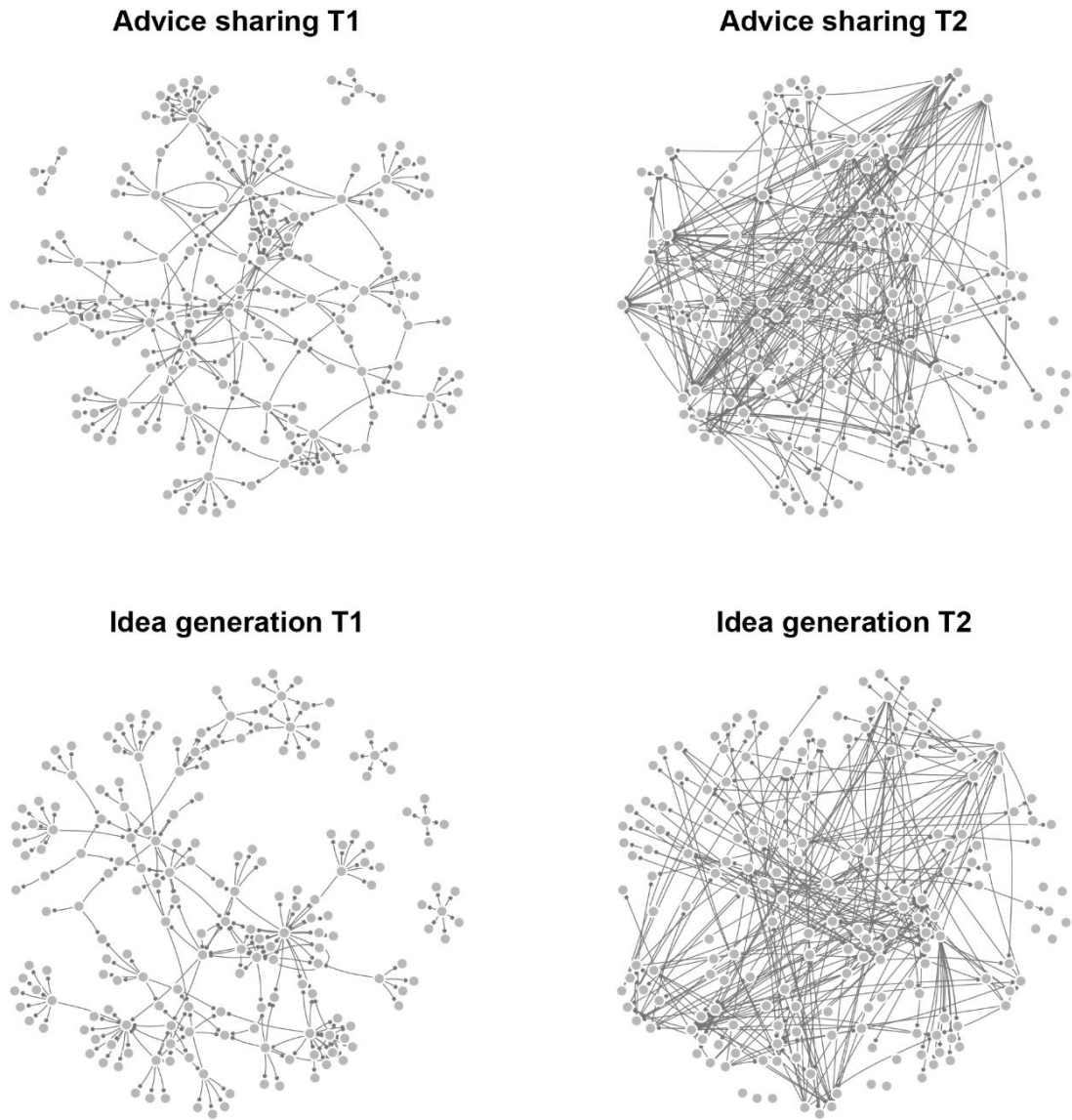

**Figure 1.** Network evolution.

The social networks representing pre-leadership connections (T1) appear significantly more fragmented, with only a few connectors/brokers. People report more interactions to seek and receive advice and fewer interactions to share ideas, as the idea generation network is slightly more clustered and less dense than the advice network. The two social networks had increased connectivity and reduced fragmentation when the course ended, as visible in Figure 1. There seems to be some degree of homophily by role, which was stronger for the advice seeking network, as participants reached out more often to colleagues holding the same roles. Concerning idea-generation, we observe more ties across roles, which implies that people share ideas with colleagues in other roles.

By looking at in-degree centrality (number of incoming ties), participants seem to go to the same 3-4 colleagues to share new ideas; by contrast, when seeking advice, they reached out to a long list of "experts".

The advice networks are relatively stable, as confirmed by the Jaccard index higher than 0.2 (Jaccard index = 0.26). Instead, the ideas sharing networks present a higher turnover (Jaccard index = 0.17), suggesting the presence of many changes between points in time. Stability is important while applying SAOMs, however, we had no problems with model convergence and estimation reliability. In fact, we could use RSiena because negative consequences for estimations can be mitigated if network changes are mostly explained by tie formation (or cessation) in the presence of a low number of outgoing ties (Ripley *et al.*, 2020). In our case, we obtained converging models, which passed all Goodness of Fit (GOF) tests. This last procedure compares the distribution of network statistics for the simulated models with the distributions of the observed networks. An adequate fit is represented by a Mahalanobis p-value larger than 0.05. We found an adequate GOF for all statistics in both advice seeking and idea sharing networks[1].

Table 2 reports the estimated results of RSiena models obtained with the Method of Moments estimation procedure. For each network, there are two models: model 1 contains the structural and exogenous effects for dyads and triads (e.g., membership division), while model 2 (full model) is more comprehensive and includes the cross-network effects. All our models converged, with individual t-statistics all being < 0.10 and the overall maximum convergence ratio being < 0.25. An effect is significant if the estimate is larger than twice the standard error in absolute value.

---

[1] GOF test results for the advice network (full model): indegree distribution, p = 0.624; outdegree distribution, p = 0.975; triads, p = 0.771; geodesic distribution p, = 0.944. GOF test results for the ideas network (full model): indegree distribution, p = 0.904; outdegree distribution, p = 0.98; triads, p = 0.295; geodesic distribution, p = 0.962.

|  | Model 1 | | | | Model 2 | | | |
|---|---|---|---|---|---|---|---|---|
|  | All convergence t ratios < 0.11. | | | | All convergence t ratios < 0.06. | | | |
|  | Overall maximum convergence ratio 0.18. | | | | Overall maximum convergence ratio 0.13. | | | |
|  | **ADVICE** | | **IDEAS** | | **ADVICE** | | **IDEAS** | |
| *Effect* | *par.* | *(s.e.)* | *par.* | *(s.e.)* | *par.* | *(s.e.)* | *par.* | *(s.e.)* |
| basic rate parameter | 10.866 | (2.503) | 16.198 | (3.386) | 9.465 | (1.306) | 14.133 | (1.883) |
| *Structural effects* | | | | | | | | |
| outdegree (density) | **–3.238*** | **(0.823)** | **–3.144*** | **(0.779)** | –2.028 | (1.276) | –1.675 | (1.783) |
| reciprocity | **1.099*** | **(0.665)** | **1.243**** | **(0.458)** | 0.431 | (0.840) | 0.928 | (0.565) |
| transitive reciprocated triplets | –1.100 | (1.694) | –0.406 | (1.465) | –4.028 | (3.782) | –0.441 | (1.677) |
| transitivity (gwesp) | **2.697*** | **(0.745)** | **1.311*** | **(0.642)** | **2.56*** | **(0.745)** | **1.114*** | **(0.529)** |
| indegree - popularity (sqrt) | **0.68**** | **(0.254)** | **0.926*** | **(0.232)** | –0.228 | (0.974) | 0.381 | (0.509) |
| outdegree - activity (centered) | –0.003 | (0.038) | –0.044 | (0.048) | 0.006 | (0.067) | –0.049 | (0.084) |
| outdegree-trunc(1) | **–6.565*** | **(1.196)** | **–6.400*** | **(0.998)** | **–8.812*** | **(2.758)** | **–8.082**** | **(3.033)** |
| anti in-isolates | **0.916*** | **(0.387)** | **0.955**** | **(0.346)** | 0.677 | (0.440) | **1.063**** | **(0.389)** |
| *Covariate effects* | | | | | | | | |
| homophily - same division | **1.398**** | **(0.234)** | **1.074**** | **(0.198)** | **1.729**** | **(0.383)** | **0.966**** | **(0.229)** |
| transitive triplets jumping by division | **–1.616†** | **(0.940)** | - | - | –1.665 | (1.356) | - | - |
| *cross-network degree effects* | | | | | | | | |
| mix - indegree popularity (sqrt) | | | | | 0.954 | (0.908) | - | - |
| mix - out-degree activity (sqrt) | | | | | 0.177 | (0.513) | - | - |
| mix - indegree popularity (sqrt) | | | | | - | - | 0.617 | (0.380) |
| mix - out-degree activity (sqrt) | | | | | - | - | –0.573 | (0.447) |
| *cross-network triadic effects* | | | | | | | | |
| effect for advice: closure of ideas | | | | | **–3.850**** | **(1.421)** | - | - |
| effect for advice: shared incoming ideas | | | | | **1.699*** | **(0.883)** | - | - |
| effect for ideas: agreement from advice | | | | | - | - | **0.349†** | **(0.188)** |

†*p* < 0.10, **p* < 0.05, ***p* < 0.01, ****p* < 0.001.

**Table 2.** RSiena models

The analysis results using SAOMs allow us to interpret whether advice seeking and idea sharing relationships reinforce each other and to what extent membership based on division influences the choice of advisors. The estimates for the Rate *(rate)* parameters are higher for the idea sharing networks. This suggests that changing partners for sharing ideas occurs more frequently than changing advisors.

Looking at Model 1, assuming that individuals have the chance to make choices, a likely scenario is that they tend to avoid creating new ties preferring instead to reciprocate their existing ties *(recip)*. The negative outdegree parameters *(outdegree - density)* indicate that advice seeking and idea sharing relationships are sparse, reflecting the balance between creating and deleting ties. Hospital professionals do not seek advice or share ideas with just anyone. Instead, this is likely to be a deliberate decision, perhaps also due to hierarchical constraints. In Model 2, reciprocity *(recip)* and density *(outdegree-density)* effects become non-significant, but their directionality remains consistent with Model 1. This could be due to the addition of cross-network effects and could suggest high interdependence between advice seeking and idea sharing relationships. Because idea sharing relationships could mediate advice relationships, considering mixed-effects provides a better explanation of endogenous tendencies.

Group formation tendencies are present in both networks and all models. Transitivity, represented by the *gwesp* parameter *(gwespFF)*, suggests that individuals tend to seek advice from the advisors of their advisors (perhaps the supervisors of their supervisors) and share ideas with the partners of the people with whom they share ideas. Although reciprocity often plays an important role in group dynamics by balancing transitivity (Block, 2015), in our networks, this effect *(transRecTrip)* is minor and non-significant, as also confirmed by the non-significant score-type test of Schweinberger (2012) ($p = 0.593$ for advice, and $p = 0.793$, for idea sharing). To help convergence, transitivity combined with reciprocity effects were retained in our models.

Regarding the degree effects, only the indegree popularity *(inPopSqrt)* is significant – in Model 1, for both networks. This shows the so-called preferential attachment tendency or Matthew effect (the *rich get richer* effect), which is the tendency for popular advisors to attract more advice seekers. Similarly, individuals who receive new ideas from many others are considered attractive and receive even more ideas. However, this effect *(inPopSqrt)* is mitigated by the interdependence of idea sharing and advice-sharing relationships, as shown in Model 2. Zero outdegree *(outTrunc(1))* and anti isolates *(antiInIso)* effects are included to help with convergence and to better account for the characteristics of our research design, with missing information about the social ties of individuals who did not attend the leadership course. In general, we find that individuals tend to share ideas or seek advice from more than one person and the majority of course participants receive at least one idea sharing or advice request.

While there is no effect of gender homophily (p = 0.182 gender advice; p = 0.456 gender ideas, following the score-type test), hospital division has a strong impact on the formation of advice seeking and idea sharing relationships. In other words, individuals tend to interact more with colleagues from the same division *(sameX)*. However, when considering triadic dynamics, a more complex picture emerges. In Model 1, we see that intermediaries do not support advice sharing across divisions (the *transitive triplets jumping by division* parameter is significant but negative, *jumpXTransTrip*). There are cases where individuals ask for advice from colleagues in their division who, in turn, ask for advice from people in other divisions. However, they do not favor the creation of bridging ties. Indeed, it is rare to have an advisor outside one's division. Nonetheless, this negative effect *(jumpXTransTrip)* decreases and becomes non-significant when considering cross-network dynamics (Model 2).

The significant and negative cross-network closure effect *(advice: closure of ideas - closure)* suggests a tendency to preserve relationships. Individuals do not share ideas with the contacts of the people from whom they ask for advice. However, there is a tendency to activate an advice link between people who receive ideas from a third common person *(advice: shared incoming ideas - sharedIn)*. For example, a person with potentially good ideas who shares them with two colleagues would favor an advice seeking relationship between the two. Similarly, having an advisor in common increases the chance to share ideas *(ideas: from advice agreement - from)*. For example, this could explain the case of two colleagues asking their supervisor for advice and then one of them going to the other to discuss and share ideas. In other terms, the transitivity effect indicates that hospital professionals are more likely to choose a colleague with whom to share ideas and be their advisor, if there is someone who suggests they do so or if someone they both know acts as a reference and intermediary.

## 5. Discussion

### *5.1. Theoretical Implications*

This study contributes to the literature in knowledge creation and advice seeking on multiple levels. First, instead of focusing on individual factors like expertise, personality or behavior of the individuals involved (Yaniv and Milyavsky, 2007), we expand the analysis to the dyadic relationship between healthcare professionals who share advice and information. Second, we assess the indirect influence of peers' relationships on individual networks, broadening the scope of the interactions that could influence the willingness to share new ideas or seek advice. By going beyond the static composition of social networks to understand dyadic factors impacting knowledge seeking and sharing, we demonstrate how individuals are indirectly influenced by the relationships of their peers. A third contribution to the literature is in terms of methodology, as we depart from traditional social science approaches, which often rely on a single snapshot of a static network and individual traits and attributes (Bucuvalas *et al.*, 2014). This study improves the understanding of the factors impacting knowledge seeking and sharing through a comparison of the structural properties of two networks before and after a professional training, that was in part designed to build connections among members, improve knowledge creation and stimulate advice seeking behaviors.

Our results suggest that hospital professionals do not seek advice or share ideas with just anyone. Instead, this is likely to be a deliberate decision, as they prefer reciprocity rather than starting new connections. The evidence that healthcare professionals prefer to reciprocate existing connections rather than establish new ones is aligned with other studies that explore the effect of homophily on emotional well-being in strong-tied networks. For example, (Elmer *et al.*, 2017) demonstrated that maintaining and cultivating existing relationships is generally less time-consuming than establishing new ones: strong ties offer people the emotional and cognitive support that weaker ties usually do not provide. Our results show the benefit of increased density in a network (a team, a unit, or an organization) which seems to encourage members to generate more diverse information. This represents an important contribution to the understanding that diagnostic decision making requires more frequent interaction among healthcare professionals from different disciplines with different skills, knowledge and connections (McKee, 1988; Ventegodt *et al.*, 2003).

The result on network density contributes to the voluminous evidence found in the area of embeddedness and network density, showing how density helps generate trust and reputation in networks, which act as governance mechanisms in social networks: the more embedded a relationship is within a cluster, the more likely bad behaviors become easily known, and the higher the chance to create a reputation cost for bad behavior (Burt, 2010; Burt *et al.*, 2013; Rivera *et al.*, 2010). As demonstrated by Aral and Van Alstyne (2011), when individuals connect with the same group of individuals (i.e. through *reciprocity*), they become aware of each other's mental models, and they are more likely to reach consensus without fully exploring the entire space of possible solutions.

It is important to consider the possibility that a tendency to only reciprocate connection, and to avoid reaching out to others, may discourage team members from developing diverse theories, decreasing the exploration of a solution space (Shore *et al.*, 2015). This may have serious implication in healthcare where the outcomes (i.e. patient care and well-being) depend on the ability to quickly integrate knowledge and find shared solutions (Zhang and Wang, 2021). Future research could adopt the lens of open innovation (Gassmann *et al.*, 2010) to explore how relying almost entirely on internal connections could negatively impact outcomes. Individuals and organizations that only rely on inflows of knowledge, rather than combining internal and external inputs, may not be successful at finding the best solution, especially when managing complex problems (Cappa *et al.*, 2022; Simeone *et al.*, 2017). This means that when individuals are embedded in dense networks and rely on the same reciprocal ties, they are likely to fall into the trap of group thinking, which is associated with decreased creativity and less innovative problem solving (Runco and Acar, 2012).

This study also indicates that transitivity (i.e., a situation where the friends of my friends are also my friends) is an important driver of tie formation and that advice and idea sharing networks shape each other. Our models confirm a *transitivity* effect, as individuals sought advice from the advisors of their advisors and shared ideas with colleagues with whom their colleagues shared ideas. This contrasts with other social network studies where healthcare members chose different colleagues when asking for advice and when sharing ideas on professional development (Burt *et al.*, 2012). The specific research context may offer an interpretation of this result: participants worked together as a cohort and immersed themselves in a leadership program that helped strengthen their strong ties and suggest to each other common advisors.

Our models also show that healthcare professionals established new connections because of a transitivity effect, which overcame the tendency to reciprocate existing ties. Results indicate that having an advisor in common increases the chance to share ideas with the same colleagues. An explanation for this comes from looking at the very nature of seeking advice. When individuals ask others for advice, they grant them prestige, showing that they respect and admire their expertise and insights (Grant, 2014). Prestige may catalyze new connections and generate a popularity effect that makes these advisors more visible outside of their immediate circle.

### 5.2. *Practical Implications*

By mapping the relational structures within teams and by looking at their evolution, hospital administrators can encourage the formation of interdisciplinary teams and promote the formation of informal ties. These can break through the hierarchical structure that typically characterizes advice-sharing dynamics in healthcare (Waring *et al.*, 2013). Not only can informal ties improve knowledge flows among hospital professionals at different levels and potentially divisions, but they can also expand network connectivity by favoring advice seeking relationships even more. Furthermore, this study suggests new ways to design and assess professional development initiatives by incorporating new indicators of success, such as the degree of inclusiveness, the transitivity factor, and the popularity effect.

Healthcare leaders must recognize the complexity of the system in which their teams operate and help them define and foster strategies to optimize team function. Given the complexity of knowledge sharing in interdisciplinary teams, hospital administrators need to organize initiatives that promote knowledge-sharing and leverage the role of professional gatekeepers (Sibbald *et al.*, 2013).

The findings could inform hospital administrators as they reflect on the nature of collaborations during professional development initiatives and promote collaboration models that may lead to positive outcomes for patients. For example, hospital leadership can encourage the creation of cross-functional teams to solve complex problems, involving external stakeholders and helping the team during the process (Ancona *et al.*, 2002). In addition, they could promote an informal exchange of ideas between professionals, which can then promote the creation of advice seeking relationships across silos. As illustrated in this study, promoting idea sharing could impact advice seeking by favoring a more balanced exchange of advice among healthcare professionals.

## 6. Conclusions

This study shows that mapping the advising network can help identify potential limitations in the flow of information among teams since we found that idea sharing and advice seeking tend to occur primarily inside rather than outside divisions. Instead of simply promoting reciprocity and connecting everybody through multiple communications technologies, which may improve coordination but decrease divergent problem solving, our results suggest that transitivity has the potential to improve diversity in the knowledge created (Aral and Van Alstyne, 2011).

Results also partially support the *rich get richer* effect, or Matthew effect (Merton, 1968), as popular advisors tend to attract more advice seekers. As suggested by research on distributed or shared leadership (Carson *et al.*, 2007; Spillane, 2005), one of the most important factors behind successful collaboration is trust-based relationships among team members. The significance of the transitivity effect in our models confirms the importance of leveraging the role of popular advisors to bridge silos and solve complex cases. Network fragmentation can be reduced if individuals who act as information brokers are incentivized to promote knowledge creation by sharing new ideas also with the advisors of their advisors. This is an important insight, as complex care requires input from and coordination with other departments to let information flow beyond divisional and departmental boundaries. Reporting relationships add complexity since team members may belong to distinct departments and many individuals belong to multiple teams.

Figure 2 summarizes the major findings of this study and highlights transitivity, homophily, and cross-networks effects.

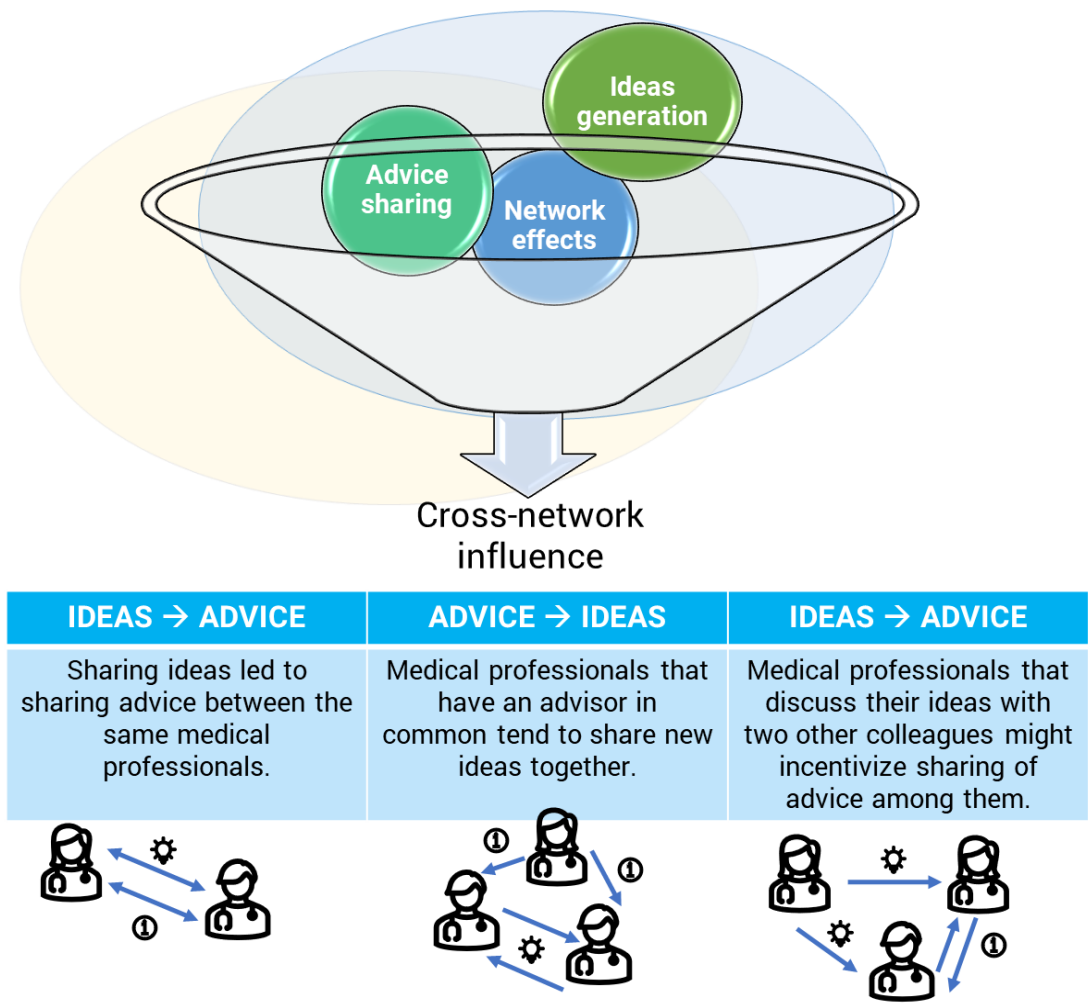

**Figure 2**. Major findings.

*6.1. Limitations and Future Research*

Despite its theoretical and practical implications, this study suffers from some limitations. The first limitation is the sample size, as we worked with a sample of 47 physicians from different medical subspecialties, representing 25% of the total number of faculty employed at the time of the study (about 200). Since the physicians involved in the study are emerging leaders, they can be considered a good representation of the faculty in leadership positions. The final network was composed of 252 individuals connected to the participants via the advice or ideas generation networks. Working with a larger sample would allow researchers to explore additional effects. A larger sample size and a more ample representation of roles would help control for individual traits such as openness to innovative ideas and extraversion that could influence interactions with peers. The second limitation is represented by the study participants, who were primarily physicians and volunteered to participate in this leadership development opportunity. Their patterns of interaction may be associated with their higher agreeableness, higher openness and extraversion, all traits that seem to impact learning attitudes and interactivity levels (Bidjerano and Dai, 2007). A third limitation is that the study participants were all in the USA, specifically New York City. Future research might analyze work dynamics in different countries to consider different cultures and see if our results are confirmed.

Future studies on the effect of professional development opportunities on knowledge sharing could involve nurses, medical assistants, pharmacists, and other roles in addition to physicians. This may help investigate the role of hierarchical relationships and cross-team interactions. This research can open new opportunities for healthcare researchers and practitioners. By applying dynamic social network analysis, researchers could explore other networks associated with highly collaborative behaviors, such as informal friendship ties (e.g., friendship) and collaborations with external partners (Zhang *et al.*, 2013). To further investigate the mechanisms that influence knowledge sharing and advice seeking, the next step is to consider the influence of additional covariates such as hierarchical position, age, tenure, and role, which have been associated with specific collaboration dynamics (Cott, 1997; Palazzolo *et al.*, 2011).

# References

Aldrich, D.P. and Meyer, M.A. (2015), "Social Capital and Community Resilience", *American Behavioral Scientist*, available at:https://doi.org/10.1177/0002764214550299.

Alrahbi, D.A., Khan, M., Gupta, S., Modgil, S. and Chiappetta Jabbour, C.J. (2022), "Challenges for developing health-care knowledge in the digital age", *Journal of Knowledge Management*, Vol. 26 No. 4, pp. 824–853.

Ancona, D., Bresman, H., Kaeufer, K. and Review, M.M. (2002), "The comparative advantage of X-teams", *MIT Sloan Management Review*, Vol. 43 No. 3, pp. 33–39.

Ancona, D., Malone, T.W., Orlikowski, W.J. and Senge, P.M. (2007), "In praise of the incomplete", *Harvard Business Review*.

Antonacci, G., Fronzetti Colladon, A., Stefanini, A. and Gloor, P. (2017), "It is rotating leaders who build the swarm: social network determinants of growth for healthcare virtual communities of practice", edited by CARAYANNIS, E. and Del Giudice, M.*Journal of Knowledge Management*, Vol. 21 No. 5, pp. 1218–1239.

Aral, S. and Van Alstyne, M. (2011), "The Diversity-Bandwidth Trade-off", *American Journal of Sociology*, Vol. 117 No. 1, pp. 90–171.

Aronson, B. (2016), "Peer influence as a potential magnifier of ADHD diagnosis", *Social Science and Medicine*, available at:https://doi.org/10.1016/j.socscimed.2016.09.010.

Bican, P.M., Guderian, C.C. and Ringbeck, A. (2017), "Managing knowledge in open innovation processes: an intellectual property perspective", *Journal of Knowledge Management*, Vol. 21 No. 6, pp. 1384–1405.

Bidjerano, T. and Dai, D.Y. (2007), "The relationship between the big-five model of personality and self-regulated learning strategies", *Learning and Individual Differences*, Elsevier BV, Vol. Barrick, M No. 1, pp. 69–81.

Block, P. (2015), "Reciprocity, transitivity, and the mysterious three-cycle", *Social Networks*, available at:https://doi.org/10.1016/j.socnet.2014.10.005.

Bucuvalas, J., Fronzetti Colladon, A., Gloor, P., Grippa, F., Horton, J. and Timme, E. (2014),

"Increasing interactions in healthcare teams through architectural interventions and interpersonal communication analysis", *Journal of Healthcare Information Management*, Vol. 28 No. 4, pp. 58–65.

Burt, R.S. (2010), *Neighbor Networks: Competitive Advantage Local and Personal*, Oxford University Press, New York, NY.

Burt, R.S., Kilduff, M. and Tasselli, S. (2013), "Social Network Analysis: Foundations and Frontiers on Advantage", *Annual Review of Psychology*, Vol. 64 No. 1, pp. 527–547.

Burt, R.S., Meltzer, D.O., Seid, M., Borgert, A., Chung, J.W., Colletti, R.B., Dellal, G., *et al.* (2012), "What's in a name generator? Choosing the right name generators for social network surveys in healthcare quality and safety research", *BMJ Quality & Safety*, Vol. 21 No. 12, pp. 992–1000.

Cangialosi, N., Odoardi, C., Battistelli, A. and Baldaccini, A. (2021), "The social side of innovation: When and why advice network centrality promotes innovative work behaviours", *Creativity and Innovation Management*, Vol. 30 No. 2, pp. 336–347.

Cappa, F. (2022), "Big data from customers and non-customers through crowdsourcing, citizen science and crowdfunding", *Journal of Knowledge Management*, Vol. 26 No. 11, pp. 308–323.

Cappa, F., Franco, S. and Rosso, F. (2022), "Citizens and cities: Leveraging citizen science and big data for sustainable urban development", *Business Strategy and the Environment*, Vol. 31 No. 2, pp. 648–667.

Carson, J.B., Tesluk, P.E. and Marrone, J.A. (2007), "Shared leadership in teams: An investigation of antecedent conditions and performance", *Academy of Management Journal*, available at:https://doi.org/10.2307/20159921.

Chesbrough, H., Vanhaverbeke, W. and West, J. (2006), *Open Innovation: Researching a New Paradigm*, Oxford University Press, Oxford, UK.

Cott, C. (1997), "'We decide, you carry it out': A social network analysis of multidisciplinary long-term care teams", *Social Science and Medicine*, available at:https://doi.org/10.1016/S0277-9536(97)00066-X.

Cross, R., Borgatti, S.P. and Parker, A. (2001), "Beyond answers: Dimensions of the advice network", *Social Networks*, available at:https://doi.org/10.1016/S0378-8733(01)00041-7.

Delva, D., Jamieson, M. and Lemieux, M. (2008), "Team effectiveness in academic primary health care teams", *Journal of Interprofessional Care*, Vol. 22 No. 6, pp. 598–611.

Elmer, T., Boda, Z. and Stadtfeld, C. (2017), "The co-evolution of emotional well-being with weak and strong friendship ties", *Network Science*, available at:https://doi.org/10.1017/nws.2017.20.

Gassmann, O., Enkel, E. and Chesbrough, H. (2010), "The future of open innovation", *R&D Management*, Vol. 40 No. 3, pp. 213–221.

Gilardi, S., Guglielmetti, C. and Pravettoni, G. (2014), "Interprofessional team dynamics and information flow management in emergency departments", *Journal of Advanced Nursing*, Vol. 70 No. 6, pp. 1299–1309.

Gino, F. (2008), "Do we listen to advice just because we paid for it? The impact of advice cost on its use", *Organizational Behavior and Human Decision Processes*, Vol. 107 No. 2, pp. 234–245.

Gino, F., Brooks, A.W. and Schweitzer, M.E. (2012), "Anxiety, advice, and the ability to discern: Feeling anxious motivates individuals to seek and use advice.", *Journal of Personality and Social Psychology*, Vol. 102 No. 3, pp. 497–512.

Granovetter, M.J. (1973), "The Strength of Weak Ties", *American Journal of Sociology*, Vol. 78 No. 6, pp. 1360–1380.

Grant, A. (2014), *Give and Take: Why Helping Others Drives Our Success*, Penguin Books, London, UK.

Griffiths, F., Cave, J., Boardman, F., Ren, J., Pawlikowska, T., Ball, R., Clarke, A., *et al.* (2012), "Social networks - The future for health care delivery", *Social Science and Medicine*, available at:https://doi.org/10.1016/j.socscimed.2012.08.023.

Grippa, F., Bucuvalas, J., Booth, A., Alessandrini, E., Fronzetti Colladon, A. and Wade, L.M. (2018), "Measuring information exchange and brokerage capacity of healthcare teams", *Management Decision*, Vol. 56 No. 10, pp. 2239–2251.

Grippa, F., Palazzolo, M., Bucuvalas, J. and Gloor, P. (2011), "Monitoring Changes in the Social Network Structure of Clinical Care Teams Resulting from Team Development Efforts", *Procedia - Social and Behavioral Sciences*.

Hofmann, D.A., Lei, Z. and Grant, A.M. (2009), "Seeking help in the shadow of doubt: The sensemaking processes underlying how nurses decide whom to ask for advice.", *Journal of Applied Psychology*, Vol. 94 No. 5, pp. 1261–1274.

Kessel, M., Kratzer, J. and Schultz, C. (2012), "Psychological Safety, Knowledge Sharing, and Creative Performance in Healthcare Teams", *Creativity and Innovation Management*, Vol. 21 No. 2, pp. 147–157.

Kilpatrick, K. (2013), "Understanding acute care nurse practitioner communication and decision-making in healthcare teams", *Journal of Clinical Nursing*, Vol. 22 No. 1–2, pp. 168–179.

Lemieux-Charles, L. and McGuire, W.L. (2006), "What do we know about health care team effectiveness? A review of the literature.", *Medical Care Research and Review : MCRR*, Vol. 63 No. 3, pp. 263–300.

Long, J.C., Cunningham, F.C. and Braithwaite, J. (2013), "Bridges, brokers and boundary spanners in collaborative networks: a systematic review", *BMC Health Services Research*, Vol. 13 No. 158, pp. 1–13.

Majchrzak, A., More, P.H.B. and Faraj, S. (2012), "Transcending Knowledge Differences in Cross-Functional Teams", *Organization Science*, Vol. 23 No. 4, pp. 951–970.

McKee, J. (1988), "Holistic health and the critique of western medicine", *Social Science and Medicine*, available at:https://doi.org/10.1016/0277-9536(88)90171-2.

Merton, R.K. (1968), "The Matthew Effect in Science: The reward and communication systems of science are considered.", *Science*, Vol. 159 No. 3810, pp. 56–63.

Mortensen, M. and Haas, M.R. (2018), "Perspective—Rethinking Teams: From Bounded Membership to Dynamic Participation", *Organization Science*, Vol. 29 No. 2, pp. 341–355.

Palazzolo, M., Grippa, F., Booth, A., Rechner, S., Bucuvalas, J. and Gloor, P. (2011), "Measuring Social Network Structure of Clinical Teams Caring for Patients with Complex Conditions", *Procedia - Social and Behavioral Sciences*, Vol. 26, pp. 17–29.

Pereira de Souza, V., Baroni, R., Choo, C.W., Castro, J.M. de and Barbosa, R.R. (2021), "Knowledge management in health care: an integrative and result-driven clinical staff management model", *Journal of Knowledge Management*, Vol. 25 No. 5, pp. 1241–1262.

Phelps, C., Heidl, R. and Wadhwa, A. (2012), "Knowledge, Networks, and Knowledge Networks: A Review and Research Agenda", *Journal of Management*, available at:https://doi.org/10.1177/0149206311432640.

Rios-Ballesteros, N. and Fuerst, S. (2021), "Exploring the enablers and microfoundations of international knowledge transfer", *Journal of Knowledge Management*, available at:https://doi.org/10.1108/JKM-04-2021-0344.

Ripley, R.M.R.M., Snijders, T.A.B.T.A.B., Preciado, P., Schweinberger, M. and Steglich, C. (2020), "Manual for RSiena version 1.2-25", *University of Oxford; Department of Statistics; Nuffield College.*

Rivera, M.T., Soderstrom, S.B. and Uzzi, B. (2010), "Dynamics of Dyads in Social Networks: Assortative, Relational, and Proximity Mechanisms", *Annual Review of Sociology*, Vol. 36 No. 1, pp. 91–115.

Runco, M.A. and Acar, S. (2012), "Divergent Thinking as an Indicator of Creative Potential", *Creativity Research Journal*, Vol. 24 No. 1, pp. 66–75.

Schweinberger, M. (2012), "Statistical modelling of network panel data: Goodness of fit", *British Journal of Mathematical and Statistical Psychology*, available at:https://doi.org/10.1111/j.2044-8317.2011.02022.x.

Seid, M., Margolis, P.A. and Opipari-Arrigan, L. (2014), "Engagement, peer production, and the learning healthcare system", *JAMA Pediatrics*, available at:https://doi.org/10.1001/jamapediatrics.2013.5063.

Shore, J., Bernstein, E. and Lazer, D. (2015), "Facts and Figuring: An Experimental Investigation of Network Structure and Performance in Information and Solution Spaces", *Organization Science*, Vol. 26 No. 5, pp. 1432–1446.

Sibbald, S.L., Nadine Wathen, C., Kothari, A. and Day, A.M.B. (2013), "Knowledge flow and exchange in interdisciplinary primary health care teams (PHCTS): An exploratory study",

*Journal of the Medical Library Association*, available at:https://doi.org/10.3163/1536-5050.101.2.008.

Simeone, L., Secundo, G. and Schiuma, G. (2017), "Knowledge translation mechanisms in open innovation: the role of design in R&D projects", *Journal of Knowledge Management*, Vol. 21 No. 6, pp. 1406–1429.

Snijders, T.A.B., van de Bunt, G.G. and Steglich, C.E.G.G. (2010), "Introduction to stochastic actor-based models for network dynamics", *Social Networks*, Vol. 32 No. 1, pp. 44–60.

Snijders, T.A.B., Faye, M. and Brailly, J. (2020), "Network dynamics with a nested node set: Sociability in seven villages in Senegal", *Statistica Neerlandica*, available at:https://doi.org/10.1111/stan.12208.

Snijders, T.A.B., Lomi, A. and Torló, V.J. (2013), "A model for the multiplex dynamics of two-mode and one-mode networks, with an application to employment preference, friendship, and advice", *Social Networks*, Vol. 35 No. 2, pp. 265–276.

Spillane, J.P. (2005), "Distributed leadership", *Educational Forum*, available at:https://doi.org/10.1080/00131720508984678.

Srinivasan, U. and Uddin, S. (2017), "A social network framework to explore healthcare collaboration", *Healthcare Ethics and Training: Concepts, Methodologies, Tools, and Applications*, available at:https://doi.org/10.4018/978-1-5225-2237-9.ch002.

Stadtfeld, C. and Pentland, A. (2015), "Partnership ties shape friendship networks: A dynamic social network study", *Social Forces*, available at:https://doi.org/10.1093/sf/sov079.

Steglich, C., Snijders, T.A.B. and Pearson, M. (2010), "Dynamic Networks And Behavior: Separating Selection From Influence", *Sociological Methodology*, available at:https://doi.org/10.1111/j.1467-9531.2010.01225.x.

Tsai, W. (2001), "Knowledge transfer in intraorganizational networks: Effects of network position and absorptive capacity on business unit innovation and performance", *Academy of Management Journal*, Vol. 44 No. 5, pp. 996–1004.

Ventegodt, S., Andersen, N.J. and Merrick, J. (2003), "Holistic medicine III: the holistic process theory of healing.", *TheScientificWorldJournal*, available

at:https://doi.org/10.1100/tsw.2003.100.

Wang, F., Srinivasan, U., Uddin, S. and Chawla, S. (2014), "Application of network analysis on healthcare", *ASONAM 2014 - Proceedings of the 2014 IEEE/ACM International Conference on Advances in Social Networks Analysis and Mining*, available at:https://doi.org/10.1109/ASONAM.2014.6921647.

Ward, V., Smith, S., House, A. and Hamer, S. (2012), "Exploring knowledge exchange: A useful framework for practice and policy", *Social Science and Medicine*, available at:https://doi.org/10.1016/j.socscimed.2011.09.021.

Waring, J., Currie, G., Crompton, A. and Bishop, S. (2013), "An exploratory study of knowledge brokering in hospital settings: Facilitating knowledge sharing and learning for patient safety?", *Social Science and Medicine*, available at:https://doi.org/10.1016/j.socscimed.2013.08.037.

Wasserman, S. and Faust, K. (1994), *Social Network Analysis: Methods and Applications*, edited by Granovetter, M., Cambridge University Press, New York, NY, available at:https://doi.org/10.1525/ae.1997.24.1.219.

Willard-Grace, R., Hessler, D., Rogers, E., Dubé, K., Bodenheimer, T. and Grumbach, K. (2014), "Team structure and culture are associated with lower burnout in primary care.", *Journal of the American Board of Family Medicine : JABFM*, Vol. 27 No. 2, pp. 229–238.

Yaniv, I. and Milyavsky, M. (2007), "Using advice from multiple sources to revise and improve judgments", *Organizational Behavior and Human Decision Processes*, Vol. 103 No. 1, pp. 104–120.

Zhang, X., Gloor, P. and Grippa, F. (2013), "Measuring creative performance of teams through dynamic semantic social network analysis", *International Journal of Organisational Design and Engineering*, Vol. 3 No. 2, pp. 165–184.

Zhang, X. and Wang, X. (2021), "Team learning in interdisciplinary research teams: antecedents and consequences", *Journal of Knowledge Management*, Vol. 25 No. 6, pp. 1429–1455.